\documentstyle[aps,twocolumn]{revtex}
\begin{document}
\draft
\title{Extracting convergent surface energies 
       from slab calculations}
\author{Vincenzo Fiorentini$^{\, 1}$
and M.  Methfessel$^{\, 2}$}
\address{
(1)\, Istituto Nazionale di Fisica della Materia and
 Dipartimento di Scienze Fisiche, Universit\`a di 
	Cagliari, via Ospedale 72, I-09124 Cagliari, Italy\\
(2)\,Institut f\"ur Halbleiterphysik, 
	P.O. Box 409, D-15204 Frankfurt(Oder), Germany}
\date{8 March 1996}
\maketitle 
\begin{abstract}    
The formation energy of a solid surface can be extracted 
from slab calculations if the bulk energy per atom is known. 
It has been pointed out previously that the resulting surface energy 
will diverge with slab thickness if the bulk energy is in error, 
in the context of calculations which used different 
methods to study the bulk and slab systems. We show here that this 
result is equally relevant for state-of-the-art computational methods
which carefully treat bulk and slab systems in the same way.
Here we compare different approaches, and present a solution to the problem
that eliminates the divergence and leads to rapidly convergent 
and accurate surface energies.
\end{abstract}

\section{INTRODUCTION} 

The knowledge of the formation energy of solid surfaces is of 
obvious importance for  surface physics and technology. Given the 
difficulties of a direct  measurement of the surface energy, accurate 
calculations \cite{met} of this quantity play a relevant role 
in surface science.

The standard method for calculating  
the surface energy $\sigma$
is to evaluate the total energy of a slab of the material 
of interest (generally with a thickness between 5 to 15 layers)
and to subtract from that the bulk energy  obtained 
from a separate calculation. This  procedure 
singles out the total energy contribution due to the presence of the 
surface. It is based on the general and intuitively appealing expression
\begin{equation}
\sigma = \lim_{N\rightarrow\infty} \frac{1}{2} (E^{N}_{\rm slab} - N\, 
 E_{\rm bulk}) \label{uno}
\end{equation}
with $E^{N}_{\rm slab}$ the total energy of a $N$-layer slab and 
$E_{\rm bulk}$ the bulk total energy; the limit is approximated in
practice by the $N$th term.
The factor of 1/2 accounts for the two surfaces of the slab.

A central but often underestimated problem with this approach 
is what value should be chosen for the bulk energy. While at first sight
this point  might be dismissed as irrelevant,
in a recent paper\cite{bot} Boettger pointed out that 
any difference between $E_{\rm bulk}$  and the 
change in $E_{\rm slab}$  with slab thickness
will cause the calculated surface energy to diverge linearly with $N$. 
Thus, increasing the slab thickness {\em must} sooner or 
later lead to unacceptable results, because the bulk energy from 
a separate calculation will never exactly equal the slope 
of the slab energy vs $N$.

In Ref.~\onlinecite{bot}, severe errors incurred by this standard 
approach were reported. Their unusual magnitude was presumably due to 
a technical matter, namely the use of two
completely different methods for calculating the bulk and surface
properties. Thus, the practical importance of the divergent behavior
of the surface energy remains unassessed for state-of-the-art methods, 
which carefully treat bulk and slab systems in the same way. The 
aim of this paper is to supply such an assessment.

In particular, the natural objection to Boettger's argument would be
that, when using the same calculational method in a technically
consistent way to obtain both 
bulk and slab quantities, this problem would simply not show up.
We show in this paper that this is not the case: the
proper choice of the bulk energy
according to Boettger's principle is an important issue in surface
energy calculations even when the bulk and slab systems are handled
consistently within accurate methods such as FP-LMTO,
pseudopotentials-plane waves, or such.
Further, we compare different approaches to remedy
the problem, presenting what seems to be the best solution.
As modern calculations advance to study more subtle
surface effects and employ thicker slabs, these results will become
increasingly relevant.

\section{RELEVANCE OF THE SURFACE ENERGY DIVERGENCE PROBLEM 
TO STATE-OF-THE ART CALCULATIONS}

In this section we present {\it ab initio} surface calculations 
demonstrating that the surface energy is not only formally, but
also practically divergent for accurate calculations. To this end, we
compare results obtained by the standard approach
with those calculated using an alternative procedure 
suggested by Boettger \cite{bot} and with a modified
approach to be described below. It will become clear that
the latter method is the most reliable by a wide margin
and should be prefered for high-accuracy applications.

\subsection{Different ways to evaluate the surface energy} 
For the standard methods, we first of all use Eq.~(1) whereby the
bulk energy was obtained from a well-converged bulk calculation
(see below).
We also considered the slightly modified version \cite{needs}
\begin{equation}
\sigma = \lim_{N\rightarrow\infty} \frac{1}{2} (E^{N}_{\rm slab} - 
\frac{N}{N_B}\, E^{N_B}_{\rm full}), \label{tre}
\end{equation}
whereby the total energies needed are that of a slab containing $N$ layers
($E^{N}_{\rm slab}$) and that of a $N_B$-atom bulk supercell
consisting of the slab plus the vacuum space between the slabs filled with
atoms ($E^{N_B}_{\rm full}$). In other words, 
the weighted energy of the filled slab
is taken as bulk energy. Using Eq.~(\ref{tre}),
many sources of difference between the slab and bulk energies
can be eliminated
since the same supercell is used for the bulk and slab systems,
albeit at the cost of an increase in computational effort.
For this approach, we used $N_B=N+7$ with N up to 11.

Boettger\cite{bot} suggested the following method to avoid
the divergence problem. For each slab thickness $N$, pick as bulk energy 
$E_{\rm bulk}$ the differential increase in the slab total energy 
upon addition of one layer of material:
\begin{equation}
\sigma = \lim_{N\rightarrow\infty} \frac{1}{2} (E^N_{\rm slab} - N\, 
\Delta E_{N}), \label{due}
\end{equation}
where $\Delta E_{N} = E^{N}_{\rm slab}-E^{N-1}_{\rm slab}$.
This formula has the obvious merit of using only slab-related 
quantities, making no reference to separately-calculated 
bulk energies. Consequently, the calculated surface energy
should not suffer from the divergence problem of Eq.~(\ref{uno}).
The price to pay is that of repeatedly calculating
total energies for slabs of increasing thickness.
At slight variance with Ref.\onlinecite{bot},
we use $\Delta E_{N} = (E^{N}_{\rm slab}-E^{N-2}_{\rm slab})/2$
in order to maintain inversion symmetry in our slabs.

As a fourth alternative, we note that as $N$ becomes large
and convergence is approached,
the definition of the surface energy in Eq.~(\ref{uno}) implies that 
\begin{equation}
E^{N}_{\rm slab} \approx 2 \sigma + N\, E_{\rm bulk}. \label{unobis}
\end{equation}
This straight-line behaviour is already dominant for very thin slabs.
This can be understood on the basis that the energy of a given atom
is determined to a large extent by its nearest-neighbor
environment \cite{Heine}.
The most straightforward way to extract the quantity $E_{\rm bulk}$ 
is to fit a straight
line to all the slab total-energy data vs.~$N$ (except for the
thinnest slabs) and to take its slope.
This value is then used in Eq.~(\ref{uno}).
This procedure uses the same data needed in Boettger's 
suggested approach. It is free of the divergence problem
because no separately-calculated bulk energy enters.
The only uncertainty in the procedure
is the assumed onset of the straight-line behaviour;
indeed, apart from the very thinnest slabs, the error bar
in $E_{\rm bulk}$ when
starting the fit at different $N$'s in the range from 
3 to 13 is $\pm$0.01 mRy (see also Table I below).

As a test case, we report results for the surface energy of Pt (001), 
evaluated by the four schemes just described.
Total energies were calculated within the local density approximation 
to density functional theory \cite{lda},
using the all-electron full-potential LMTO method \cite{fp}.
Slab thicknesses of up to 15 layers, and  a vacuum spacing 
equivalent to seven bulk layers were used. The
slabs were left unrelaxed in the ideal fcc geometry. 
The technical ingredients (basis set, k-point 
summation, etc.) are given in Ref. \onlinecite{met}.

\subsection{Results and discussion} 

Figure \ref{f-uno} shows the calculated surface energies as function
of slab thickness when the four described methods are used.
The two approaches using a separately-calculated bulk energy 
[Eqs.~(\ref{uno}) and (\ref{tre})] are shown by open circles and 
triangles, respectively. Both evidently suffer from the divergence 
problem, shown by the linear decrease as the slabs are made thicker. 
Boettger's approach (taking the bulk energy as differential increase
of the slab energy, filled diamonds) does indeed give a surface energy 
which does not have this systematic linear behaviour. 
Unfortunately, it shows large oscillations which decrease only very slowly
as the slab is made thicker.
If only these three techniques were available, the best legitimate 
conclusion would be that the surface energy lies somewhere 
between 1.21 and 1.26 eV/atom. This is an uncertainty of 5\% even though 
slabs as thick as 15 layers were considered.

The fourth technique (fitting a straight line to the 
$E^{N}_{\rm slab}$ data to obtain $E_{\rm bulk}$) is shown by filled
squares. In comparison to the other approaches, very fast convergence to
a stable value is achieved. 
We can now accurately determine the calculated surface energy 
to be 1.246 eV, a value which is  numerically stable to
within 0.5 meV. This reduces the uncertainty to below 0.1\%.
In order to suppress the weak residual oscillation, we averaged over 
the last three points in the curve: however, the deviations from the 
average are below 0.5 meV.
(For completeness, in Table I we list the raw data 
for the slab total energies,
where a constant offset of  --36800 Ry/atom was subtracted for convenience,
and the bulk energies obtained by linear 
fittings to $E^{N}_{\rm slab}$ vs. $N$ starting at different  
values of the slab thickness $N$.)

Our main point here is that one should be wary of 
the standard technique (which uses a separately calculated bulk energy)
even for calculations of high accuracy. We can compare the
bulk energy as deduced from the slope of the $E^{N}_{\rm slab}$ data
($-36806.84240$~Ry) with that from the
well-converged bulk crystal calculation ($-36806.84182$~Ry),
finding a difference of only 0.6~mRy~$\simeq 0.01$~eV.
Despite this very small discrepancy, the undesired linear behaviour 
in the calculated surface energy is already prominent for thicknesses
of eight or more layers. The accumulated bulk error for 
the (typically used) slab thickness of around seven
is already uncomfortably large, namely 4.2~mRy $\simeq 0.06$~eV.
Optimistically going to thicker slabs would soon lead to 
unacceptable values of the surface energy. Although previously 
calculated surface energies using the standard technique 
with slab thicknesses of below ten layers can be considered
reasonably reliable, it is clearly important to keep  
the problem addressed here in mind when doing surface calculations.

To some extent, the severity of the problem will depend on the
calculational method used. In terms of k-points and 
selfconsistency iterations, our bulk crystal energy
was converged to within 0.01 mRy. Thus, the 
discrepancy in $E_{\rm bulk}$ presumably comes from the k-point mesh 
for the slab, which consisted of 15 irreducible special points in 
the $xy$ plane. It is highly desirable to be able to use a mesh
of this typical size, independent of the exact {\it ab-initio}
scheme used. In this context, our conclusions apply in exactly
the same way to other methods.

Although our suggested scheme at first sight looks like a mere
numerical procedure, there is a clear theoretical background to it. 
The squares in Fig.~1 show small but definite oscillations
of the surface energy as function of the slab thickness with
a period of about six to eight layers. These quantum size effects are 
due to the finite thickness of the slab.
The problem of Boettger's
scheme is that it artificially magnifies these oscillations 
by a large factor because the bulk energy is calculated from
two slabs of similar thickness. A technique which exploits the
overall linear behaviour of the $E^{N}_{\rm slab}$ data, such as ours,
eliminates this problem.

Finally, we point out that our tests up to now, while informative,
used all the $E^{N}_{\rm slab}$ data up to $N$=15 to obtain the
converged surface energy. In practice, the aim is to
use the data from slabs up to a thickness of typically 7 to 9 layers.
Using the same procedure as before, this gives surface energies
of 1.2456 and 1.2457 eV/atom, respectively. These values are much closer
to the converged value than those of the three competing approaches.
For completeness, we mention that analogous results have been 
obtained for Al (001)\cite{doni}, and Rh and Ir low-index 
faces \cite{rhod}, as will be presented elsewhere.

\section{SUMMARY}  

In summary, 
it has been previously pointed out\cite{bot} that the calculated 
surface energy 
will diverge with slab thickness if a bulk energy is used which
is not exactly equal to the slope of the slab energy 
vs. slab thickness. Here, we have investigated this phenomenon 
in the context of accurate state-of-the art computational methods
which are careful to treat bulk and slab systems in the same way.
The results show that the  effect must be taken seriously
for this type of calculation also. The problem can be easily solved
by obtaining the bulk crystal energy directly as the slope of the
$E^{N}_{\rm slab}$ data, but care should be
taken to eliminate quantum size effects. This can be done
by making an overall linear fit to the slab total energy
as function of the thickness.

  
\begin{figure}
\caption{Calculated surface energy for Pt (001) as a function of
 slab thickness (see text for symbol explanation).}
\label{f-uno}
\end{figure}

\begin{table}[ht]
\begin{center}
\begin{tabular}{|c|rr|}
\multicolumn{1}{c}{$N$} &
\multicolumn{1}{l}{$E^{N}_{\rm slab}$} &
\multicolumn{1}{l}{$E^{N}_{\rm bulk}$}  \\
\hline
1  &  -6.65451& -6.8426  \\
3  &  -20.34388 & -6.8424   \\
5  &   -34.02959 & -6.8424   \\
7  &  -47.71315  & -6.8425   \\
9  &  -61.39835 & -6.8424   \\
11  &  -75.08339 & -6.8424   \\
13 &  -88.76793  & -6.8425   \\
15 &  -102.45292  & ---  \\
\hline
bulk & --- & -6.8418 \\
\end{tabular}
\end{center}\caption[T3]{\footnotesize 
In columns from left to right:
total energies per atom in Pt (100) slabs of thickness $N$
after subtraction of $N\times$36800 Ry (center);
bulk energy as the slope extracted from linear fitting 
of
slab total energies vs. $N$, starting the fit at different values of 
$N$.}\label{t3}
\end{table}
\end{document}